\documentclass[preprint,
showpacs,amsmath
,nofootinbib
]{revtex4-1}
\usepackage{graphicx}
\usepackage{dcolumn}
\usepackage{bm}
\usepackage{epstopdf, epsf}
\usepackage{hyperref}
\everymath{\displaystyle}
\begin{document}
\title{Curiosity about the dust matter in the cosmological context}
\author{Amir Ghalee}
\email[\,Electronic address:\ ]{ghalee@ut.ac.ir}
\affiliation{{Department of Physics, Tafresh  University,
P. O. Box 39518-79611, Tafresh, Iran}}
\begin{abstract}
We propose a model for the dust matter in the cosmological context. The model contains a scalar field with a kinetic term nonminimally coupled to gravity.
By investigating the background and perturbative equations, it is demonstrated that the scalar field has the same dynamics as the dust matter.
We have also considered the cosmological constant in the model. It turns out that the model has not exotic behavior. Thus, a universe including the scalar field and the cosmological constant evolves just as our Universe. Moreover, we have
added the quadratic term in the action. It is shown that the quadratic term can be ruled out by its consequences.
\pacs{98.80.Cq}
\end{abstract}
\pacs{98.80.Cq}
\maketitle
\section{\label{sec:level1}INTRODUCTION}
In the Newtonian cosmology the dust matter is defined as ``matter whose pressure, $p$, is negligible compare to its energy density $\rho$ " \cite{mukhanov}.\\
In the relativistic cosmology, the stress-energy tensor, $T_{\mu\nu}$, is parametrized as an effective fluid( Appendix B of Ref. \cite{weinberg}),
\begin{equation}\label{0-1}
 T^{\mu}_{\nu}=pg^{\mu}_{\nu}+(p+\rho)u^{\mu}u_{\nu}+\triangle T^{\mu}_{\nu},
\end{equation}
where $u^{\mu}$ is ``the velocity vector " and the imperfect part of the fluid is represented by $\triangle T^{\mu}_{\nu}$.
In almost all cases, it is
possible to construct( or reconstruct) the stress-energy tensor by variation of the matter action, $S_{M}$, with respect to the metric.\\
It is convenient to describe the dynamics of matter with relations between components
of the stress-energy tensor. For example, consider a scalar field minimally
coupling to the gravity, which is described by an action
\begin{equation}\label{0-2}
S_{\phi}=\int d^{4}x\sqrt{-g}\left[-\frac{1}{2}g^{\mu\nu}\partial_{\mu}\phi\partial_{\nu}\phi-V(\phi)
\right],
\end{equation}
in which the Friedmann-Robertson-Walker (FRW) background metric is adopted. For $V(\phi)=\frac{1}{2}m^{2}\phi^{2}$, we have two
phases for the scalar field, which have important consequences for the dynamics of the universe. The first is the so-called inflationary phase and can be characterized as $\rho_{\phi}\approx-p_{\phi}$.
The other phase can be described as $p_{\phi}\approx0$.
Since the later phase occurs when the scalar field oscillates around the minimum of its potential, it is
the attractor phase. At this phase the scalar field behaves like the dust matter.
For the quadratic potential in \eqref{0-2}, it has been shown that the dustlike phase can be represented as\cite{mukhanov}
\begin{equation}\label{0-4}
H(t)=\frac{2}{3t}\quad,\quad \phi(t)=\frac{\sqrt{6}H(t)}{\kappa m}\cos\left(mt\right),
\end{equation}
where higher order terms have been neglected \cite{mukhanov}. The Hubble parameter in \eqref{0-4} shows that the Universe is driven to the dustlike
phase \cite{sawicki}.\\
Obviously we have not a term in \eqref{0-2} to regard it as \textit{the dust matter term}, and the dust phase is the consequence of dynamics of the scalar field.
It is worth noting that almost all other matters of the Universe, e.g. photons, have their action( or effective action), but it seems that the
dust matter is an exception.\\
In this work we present a novel scenario to obtain the dustlike phase with the following action
\begin{equation}\label{0-5}
S=\int d^{4}x\sqrt{-g}\left[\frac{R}{2\kappa^{2}}+\frac{1}{2}\alpha^{2}G^{\mu\nu}\partial_{\mu}\varphi\partial_{\nu}\varphi-V(\varphi)
\right],
\end{equation}
where $G^{\mu\nu}$ is the Einstein's tensor and $\kappa^{2}=8\pi G$ \cite{list}.\\
At first glance, the case in which $V(\varphi)=0$ is not interesting. On the contrary, we want to show
that this case is not as dull as it seems. Indeed, the main aim of this work is to show that the second term in the action \eqref{0-5} can be
interpreted as an effective field description of the dust matter in the cosmological context.\\
With the stated perspective, two other cases, $V(\varphi)=\Lambda$ and the quadratic term, will be studied and
their results will be interpreted.\\
The organization of this paper is as follows: in Sec. II we briefly review the properties of the dust matter in the Newtonian theory
and general relativity. The background equations and general cosmological perturbation equations are
given in Sec. III. In Sec. IV the case in which $V(\varphi)=0$, ``pure dust, " is studied and we show that why the second term in \eqref{0-5} can be interpreted as the dust matter.
Sections V and VI are devoted to discuss the other potentials. We summarize our findings and discuss the results in Sec. VII. In the Appendix the
second order action for the scalar metric perturbations in which $V(\varphi)\neq0$ is given.
\section{\label{sec:level1}Summary of what we have known}
In this section the well-known results for the behavior of dust matter, in the cosmological context, are presented \cite{weinberg,mukhanov}, then the results are converted to the comoving gauge, which is used in this work.\\
In the Newtonian theory, it is possible to construct an expanding universe with nonrelativistic matter \cite{mukhanov}.
It turns out that the background energy density, $\overline{\rho}$, scales as $t^{-2}$, and the fractional amplitude of the density perturbation, $\delta\equiv\delta\rho/\overline{\rho}$, is (ignoring decaying modes)
\begin{equation}\label{1-1}
\delta^{Newtonian}=A(x)t^{2/3},
\end{equation}
where $A(x)$ is a time-independent function.
Since the Newtonian theory is a very good approximation for deep inside the horizon, any model for the dust matter must have the same behavior in this limit.\\
In Einstein's gravity, with the flat FRW background metric, the dust matter is \textit{defined} as matter with $\overline{p}=0$. The
background energy density scales as $a^{-3}$, where $a$ is the scale factor. Since for the dust matter dominated era we have $a^{3}H^{2}=$constant, it turns out that $\overline{\rho}=3\kappa^{-2} H^{2}$ \cite{mukhanov}. Moreover, usually, it is \textit{assumed}
that the anisotropic stress terms in the stress-energy tensor are absent, $\triangle T^{\mu}_{\nu}=0$, otherwise, such terms must be defined separately and put in $T_{\mu\nu}$.\\
Ignoring decaying modes, the scalar metric perturbations are given by \footnote{Henceforth, the Laplacian is replaced with $-k^{2}$ in the final results, where $k$ is the wavenumber.} \cite{mukhanov}
\begin{equation}\label{1-2}
\delta^{(N)}_{out}\simeq C(x),\hspace{2pt}  \delta^{(N)}_{in}\simeq k^{2}aC(x),\hspace{2pt}\delta^{(N)}\hspace{2pt} p=0,\Pi^{S(N)}=0,
\end{equation}
where $(N)$ indicates the Newtonian gauge and the subscripts $in$ and $out$ stand for inside the horizon, $k^{2}a\ll1$, and outside the horizon, $k^{2}a\gg1$, respectively. $\Pi^{S}$ is the scalar part of the anisotropy inertia term, which appears in $\delta T_{ij}$ as $a^{2}\partial_{i}\partial_{j}\Pi^{S}$.\\
One can convert the results from the Newtonian gauge to the comoving gauge \cite{weinberg,mukhanov}.
The density perturbation,  by ignoring decaying modes, takes the following form
\begin{equation}\label{1-3}
  \delta=k^{2}aC(x),
\end{equation}
both inside and outside the horizon.\\
It is \textit{assumed} that $\delta^{(N)}p=0$, so, it follows that $\delta p=0$. But in the comoving gauge, anisotropic stress does not vanish and has
the following decaying solution
\begin{equation}\label{1-4}
 \partial^{2}\Pi^{S}=\mathcal{O}(t^{\frac{-11}{3}})\Rightarrow \frac{\partial^{2}\Pi^{S}}{\overline{\rho}}=\mathcal{O}(t^{\frac{-5}{3}}),
\end{equation}
where $\partial^{2}$ is the Laplacian in 3D Euclidian space. Since $\overline{p}=0$, $\overline{\rho}$ is used to make a
dimensionless quantity and compare the result with the background quantities.\\
The vector metric perturbations are suppressed both in the Newtonian gauge and the comoving gauge. In this work the scalar field is considered in the comoving gauge without
any coupling to a vector field; thus, the vector metric perturbations of our models vanish.\\
As for the tensor metric perturbation, $\gamma_{ij}$, using the following expansion,
\begin{equation}\label{1-5}
 \gamma_{ij}=\int\frac{d^{3}k}{(2\pi)^{3/2}}\sum_{s=\pm}\epsilon_{ij}^{s}(k)\gamma_{k}^{s}(t)e^{i\overrightarrow{k}.\overrightarrow{x}},
\end{equation}
 where $\epsilon_{ii}=k^{i}\epsilon_{ij}=0$ and $\epsilon_{ij}^s(k)\epsilon_{ij}^{{s'}}(k)=2\delta_{s'}$, it follows that\cite{mukhanov}
 \begin{equation}\label{1-6}
 \gamma_{k}^{s}(t)=B\frac{\sin\sqrt{\theta}-\sqrt{\theta}\cos\sqrt{\theta}}{\theta^{\frac{3}{2}}}+D\frac{\cos\sqrt{\theta}+\sqrt{\theta}\sin\sqrt{\theta}}{\theta^{\frac{3}{2}}}
\end{equation}
where, $\theta\equiv9k^{2}a$, and $B,D$ are integration constant. Therefore
\begin{equation}\label{1-7}
  \gamma_{k}^{s}(t)|_{out}\approx B,\quad \gamma_{k}^{s}(t)|_{in}\approx \frac{\exp[{\pm i\sqrt{\theta}}]}{\theta}
\end{equation}
In the next sections, we compare our results with the stated results in this section. In this work we set $a^{3}H^{2}=1$ for the dust dominated era.
\section{\label{sec:level1}General equations}
To obtain equations of motion, the ADM formalism is used in which a metric can be written as \cite{wald}
\begin{equation}\label{2-1}%
ds^{2}=-N^{2}dt^{2}+h_{ij}(dx^{i}+N^{i}dt)(dx^{j}+N^{j}dt),
\end{equation}
where $N$ and $N^{i}$ are the laps and shift, respectively.\\
The background metric is the flat FRW metric,
\begin{equation}\label{2-metricback}%
ds^{2}=-N(t)dt^{2}+a(t)^{2}\delta_{ij}dx^{i}dx^{j}.
\end{equation}
In the comoving gauge, the scalar metric perturbations, $\zeta$, and the tensor metric perturbations ,$\gamma_{ij}$, are defined as
\begin{equation}\label{2-2}
  \delta\varphi=0\quad h_{ij}=a^{2}[(1+2\zeta)\delta_{ij}+\gamma_{ij}],
\end{equation}
where $\partial_{i}\gamma_{ij}=0$, $\gamma^{ii}=0$.\\
Following \cite{maldacena}, plugging Eqs. \eqref{2-1} and \eqref{2-2} into the action \eqref{0-5}, the following action obtained:
\begin{eqnarray}\label{2-3}
S=\frac{1}{2\kappa^{2}}\int dta^{3}\sqrt{h}N[^{(3)}&&R(1+\frac{\kappa^{2}\alpha^{2}\dot{\varphi}^{2}}{2N^{2}})+(K_{ij}K^{ij}-K^{2})\nonumber\\
&&\times(1-\frac{\kappa^{2}\alpha^{2}\dot{\varphi}^{2}}{2N^{2}})-2\kappa^{2}V(\varphi)],
\end{eqnarray}
where $^{(3)}R$ is the three-dimensional curvature and the extrinsic curvature, $K_{ij}$, is defined as
\begin{equation}\label{2-4}
 K_{ij}=\frac{1}{2N}(\dot{h}_{ij}-2\nabla_{(i}N_{j)}),
\end{equation}
where $\nabla_{i}$ is the spatial covariant derivative.\\
Also we used the following decomposition of the Einstein's tensor in the ADM formalism \cite{wald}:
\begin{equation}\label{ref}
2G_{\mu\nu} n^{\mu}n^{\nu}=^{(3)}R-(K_{ij}K^{ij}-K^{2})
\end{equation}
and consider $\dot{\varphi}$ as normal to the spacelike hypersurface.\\
By varying the action \eqref{2-3} with respect to the laps and shift, it turns out that
\begin{subequations}
\begin{align}
        &\nabla_{i}\left[(K^{ij}-h^{ij}K)(1-\frac{\kappa^{2}\alpha^{2}\dot{\varphi}^{2}}{2N^{2}})\right]=0\label{2-5-1},\\
        &^{(3)}R(1-\frac{\kappa^{2}\alpha^{2}\dot{\varphi}^{2}}{2N^{2}})-(K_{ij}K^{ij}-K^{2})(1-\frac{3\kappa^{2}\alpha^{2}\dot{\varphi}^{2}}{2N^{2}})-
        2\kappa^{2}V(\varphi)=0 \label{2-5-2}.
\end{align}
\end{subequations}
To solve the above equations, we expand the shift and laps as \cite{maldacena}
\begin{equation}\label{2-6}
N=N^{(0)}+N^{(1)}+\cdots,\quad \quad N_{i}=N_{i}^{(0)}+N_{i}^{(1)}+\cdots
\end{equation}
where $N^{(0)}=1$ and $N_{i}^{(0)}=0$. The shift can be decomposed into a scalar field and a vector field as
\begin{equation}\label{2-7}
  N_{i}^{(n)}=\partial_{i}\psi+N_{i}^{T}\quad,\quad \partial_{i}N_{i}^{T}=0.
\end{equation}
For the background metric, Eq. \eqref{2-5-1} is trivial and Eq. \eqref{2-5-2} results in
\begin{equation}\label{2-fri}
 H^{2}=\frac{\kappa^{2}}{6}\left[\dot{\varphi}^{2}9\alpha^{2}H^{2}+2V(\varphi)\right]
\end{equation}
Therefore, when $V(\varphi)=0$ we can define the following energy density for the model
\begin{equation}\label{energy-den}
\bar{\rho}_{d}\equiv\frac{\dot{\varphi}^{2}9\alpha^{2}H^{2}}{2}.
\end{equation}
Also, the equation of motion for the scalar field, $\varphi$, is obtained by variational with respect to $\varphi$ and setting $N=1$, $N^{i}=0$, as
\begin{equation}\label{2-8}
3\alpha^{2}H^{2}\ddot{\varphi}+6\dot{\varphi}H\dot{H}\alpha^{2}+9\alpha^{2}H^{3}\dot{\varphi}=-\frac{dV(\varphi)}{d\varphi}.
\end{equation}
In general context, the equation of motion for the scalar field is
\begin{equation}\label{2-8-general}
\alpha^{2}\nabla_{\mu}(G^{\mu\nu}\nabla_{\nu}\varphi)=\frac{dV(\varphi)}{d\varphi},
\end{equation}
where the Bianchi identity is used. Since for the maximally symmetric space time $G_{\mu\nu}$ is proportional to $g_{\mu\nu}$, the scalar field
evolves as if the scalar field with the normal kinetic term; in such backgrounds. Also if $dV(\varphi)/d\varphi=0$, the above equation
is trivial for the vacuum solution of the Einstein equation.\\
For the first order perturbation, Eq. \eqref{2-5-1} gives
\begin{equation}\label{2-p-1}
N^{(1)}=\frac{\dot{\zeta}}{H},
\end{equation}
and from Eq. \eqref{2-5-2} it follows that
\begin{equation} \label{2-p-2}
(3\alpha^{2}\kappa^{2}\dot{\varphi}^{2}-2)H\partial^{2}\psi=(2-\alpha^{2}\kappa^{2}\dot{\varphi}^{2})\partial^{2}\zeta
-9\alpha^{2}\kappa^{2}\dot{\varphi}^{2}Ha^{2}\dot{\zeta}.
\end{equation}
To find the second-order action for the scalar metric perturbation, $h^{S}_{ij}\equiv 2a^{2}\delta_{ij}\zeta$, following\cite{maldacena}; replace the shift and laps in \eqref{2-3},
and then use the background solutions. The stated procedure, with some integration by parts, leads us to the following result
\begin{widetext}
\begin{equation}\label{2-seond-scalar}
S^{S}=\frac{1}{2\kappa^{2}}\int dta\left[\left(2\frac{\dot{H}}{H^{2}}(\alpha^{2}\kappa^{2}\dot{\varphi}^{2}-2)-5\alpha^{2}\kappa^{2}\dot{\varphi}^{2}\right)\zeta\partial^{2}\zeta+3\alpha^{2}\kappa^{2}\dot{\varphi}^{2}a^{2}\dot{\zeta}^{2}
+4\alpha^{2}\kappa^{2}\dot{\varphi}^{2}\dot{\zeta}\partial^{2}\psi\right]
\end{equation}
\end{widetext}
For the cases that we will study, the Hubble parameter is obtained as an explicit function of time. So, in Eq. \eqref{2-seond-scalar},
$\dot{H}$ is kept.\\
For $V(\varphi)\neq0$, $\psi$ can be eliminated from the above action by using \eqref{2-p-2}, the result is given in the Appendix. As for $V(\varphi)=0$, pure dust, see the next section.\\
Finally, since the kinetic term is nonminimally coupled to the tensor field, $G_{\mu\nu}$, the quest for any change in the spectrum of tensor metric perturbations must be investigated.\\
To find the tensor metric perturbation, $h^{T}_{ij}\equiv a^{2}\gamma_{ij}$, inserting $h^{T}_{ij}$ in the action \eqref{2-3}, and keep only terms of second order, the result is
\begin{eqnarray}\label{3-19}
S^{T}=\frac{1}{8}\int dtd^{3}x&&[(1+\frac{\alpha^{2}\kappa^{2}{\dot{\varphi}}^{2}}{2})a\gamma_{ij}\partial^{2}\gamma_{ij}\nonumber\\
&&+(1-\frac{\alpha^{2}\kappa^{2}{\dot{\varphi}}^{2}}{2})a^{3}\dot{\gamma_{ij}}\dot{\gamma_{ij}}].
\end{eqnarray}
\section{\label{sec:level1} the	pure dust matter}
Now, we focus on the second term in \eqref{0-5} by taking $V(\varphi)=0$. We will show that it can be interpreted as the dust
matter term.\\
The following equations are given by \eqref{2-fri} and \eqref{2-8}:
\begin{equation}\label{3-1}
\begin{split}
& H^{2}=\frac{3\alpha^{2}}{2}H^2\dot{\varphi}^{2} ,\cr
&\ddot{\varphi}(3H^{2}\alpha^{2})+3H\dot{\varphi}(3H^{2}\alpha^{2})+6\dot{\varphi}H\dot{H}\alpha^{2}=0.
\end{split}
\end{equation}
These equations can be solved as
\begin{equation}\label{3-2}
\dot{ \varphi}^{2}=\frac{2}{3\kappa^{2}\alpha^{2}},\quad H^{2}a^{3}=const.\equiv1,
\end{equation}
also, from \eqref{energy-den}, the energy density scales as $a^{-3}$. Thus, for this case, the Universe is driven by the ``dust matter". But, whether the perturbations have similar behavior to the dust matter must be investigated.
\subsection{\label{sec:level1}	The scalar metric perturbations}
From \eqref{3-2} it turns out that the left-hand side of Eq. \eqref{2-5-2} is vanished, so for this case
\begin{equation}\label{3-3}
N^{(1)}=\frac{\dot{\zeta}}{H}\quad,\quad \partial^{2}\zeta=\frac{9}{2}Ha^{2}\dot{\zeta}.
\end{equation}
In comparison with other models, it is unusual that $\psi$ is not present in the constraint equations. We will get back to this point.\\
Following \cite{maldacena}, to use the second order action, inserting the background solutions and the constraint equations, Eq. \eqref{3-3} in \eqref{2-seond-scalar}
and then integrating by parts. By this procedure one can convert terms like $\dot{\zeta}^{2}$ in \eqref{2-seond-scalar}
to $\dot{\zeta}\partial^{2}\zeta$ and then to $\zeta\partial^{2}\zeta$. The procedure gives the second-order action as
\begin{equation}\label{3-6}
S_{2}=\frac{1}{9\kappa^{2}}\int dtd^{3}x\left[\frac{1}{2}a\zeta\partial^{2}\zeta+12a\dot{\zeta}\partial^{2}\psi\right].
\end{equation}
One can use \eqref{3-3} to eliminate $\dot{\zeta}$, but for what we want to do, the above form for $S_{2}$ is sufficient.\\
Recall that $\psi$ acts as Lagrange multiplier in the ADM formalism. So variation with respect to it gives  the constraint
equation. The fact that we cannot remove it from the above action, as $V(\varphi)\neq0$ cases, shows
that we have an additional constraint in the model for pure dust matter.\\
From \eqref{3-6}, variation with respect to $\psi$ gives
\begin{equation}\label{3-7}
\partial^{2}\dot{\zeta}=0 \Rightarrow\partial^{2}\zeta=0.
\end{equation}
Using Eq. \eqref{3-3} and the above result, we have
\begin{equation}\label{3-8}
\dot{\zeta}=0.
\end{equation}
Substituting the above result in \eqref{3-6} and variation with respect to $\zeta$ does not give additional information.\\
But $\psi$ is not determined at this step.
To find it, we first calculate the density perturbation in the comoving gauge, then we compare it with \eqref{1-1}, i.e. to the result of the Newtonian theory.\\
In the comoving gauge, for the pure dust matter we have\cite{weinberg}
\begin{equation}\label{3-9}
\delta=-\frac{2a}{3}\partial^{2}[\zeta+H\psi].
\end{equation}
Using Eq. \eqref{3-7} it follows that
\begin{equation}\label{3-10}
 \delta=-\frac{2}{3}aH\partial^{2}\psi.
\end{equation}
From the Newtonian result \eqref{1-1}, we conclude that
\begin{equation}\label{3-11}
\psi=\frac{X(x)}{H},
\end{equation}
where $X(x)$ is an arbitrary function of space. Therefore, after absorption of $2/3$ in $X(x)$, we have
\begin{equation}\label{3-133}
\delta=k^{2}X(x)a.
\end{equation}
which is in agreement with \eqref{1-3}. It is interesting that we just use the Newtonian results to find $\psi$, but the
final result is in agreement with \eqref{1-3}.\\
In the comoving gauge the scalar part of anisotropic stress is obtained as \cite{weinberg}
\begin{equation}\label{3-13}
  \kappa^{2}a^{2}\partial^{2}\Pi^{S}=a^{2}(\partial_{t}+3H)\frac{\partial^{2}\psi}{a^{2}}+\partial^{2}(\zeta+\frac{\dot{\zeta}}{H}).
\end{equation}
Combining Eqs. \eqref{3-7} and \eqref{3-11} with \eqref{3-13} results in
\begin{equation}\label{3-14}
   \kappa^{2}a^{2}\partial^{2}\Pi^{S}=k^{2}\frac{5}{2}X(x).
\end{equation}
Hence, similar to \eqref{1-4}, the scalar part of anisotropic stress is decaying function of time. But
comparison of Eqs. \eqref{1-4} and \eqref{3-14} shows that it decays more slowly in this model.
To clarify this result, let us use \eqref{3-14}, $\bar{\rho}\simeq a^{-3}$ and $\delta\approx a$, to write the following statements
\begin{subequations}
\begin{align}
        &\frac{\partial^{2}\Pi^{S}}{\bar{\rho}}\approx a\label{3-143a} ,\\
        &\frac{\partial^{2}\Pi^{S}}{\delta\rho}= const.\label{3-143b}
\end{align}
\end{subequations}
The relation \eqref{3-143a} indicates that the anisotropic stress is growing compared with $\bar{\rho}$. Also, \eqref{3-143b} shows
that the rate of structure formation and anisotropic stress are the same.\\
One can find $\delta p$ by the following formula ( in the comoving gauge and for $\bar{p}=0$):
\begin{equation}\label{3-16}
\frac{\delta p}{\overline{\rho}}=-\frac{\dot{\zeta}}{H}-\frac{2}{3}\frac{\partial^{2}\Pi}{\overline{\rho}}.
\end{equation}
From \eqref{3-14} we have
\begin{equation}\label{3-18}
   \frac{\delta p}{\overline{\rho}}\approx k^{2}X(x)a.
\end{equation}
But, since $\overline{\rho}\simeq a^{-3}$, it follows that $\delta p\approx \mathcal{O}(a^{-2})$.\\
Hence, although in this model $\delta p$ is not exactly vanished, it is a decaying function of time.
\subsection{\label{sec:level1}	The tensor metric perturbations}
Using Eqs. \eqref{3-19}, \eqref{3-2} and then $x\rightarrow\sqrt{2}x$ results in
\begin{equation}\label{3-20}
S^{T}=\frac{\sqrt{2}}{6\kappa^{2}}\int dtd^{3}x\left[a\gamma_{ij}\partial^{2}\gamma_{ij}+a^{3}\dot{\gamma_{ij}}\dot{\gamma_{ij}}\right].
\end{equation}
The above action, aside from an overall number, is the same as the action for the gravitational waves in the Hilbert-Einstein action.
So dynamics of the gravitational waves, which is given by variation of $S^{T}$ with respect to $\gamma_{ij}$ as
\begin{equation}\label{3-21}
 \partial^{2}\gamma_{ij}=3Ha^{2}\dot{\gamma_{ij}}+a^{2}\ddot{\gamma_{ij}}.
\end{equation}
has the same solutions as \eqref{1-6}.\\
Note that, since the tensor part of anisotropic stress, $\Pi^{T}_{ij}$, is defined as \cite{weinberg}
\begin{equation}\label{3-22}
16\pi Ga^{3}\Pi^{T}_{ij}=3a^{2}\dot{a}\dot{\gamma}_{ij}+a^{3}\ddot{\gamma_{ij}}-a\partial^{2}\gamma_{ij},
\end{equation}
therefore, for this case $\Pi^{T}_{ij}=0$. As we will see, present of other terms in \eqref{0-5}, may be changed this result.
\section{\label{sec:level1}	implications of the cosmological constant term}
Observations of the cosmic microwave background and the large scale structure are consistent with
an accelerated expansion phase at present time, which follows after the dust matter dominated era \cite{plank-data}.
The accelerated expansion phase can be explained by the cosmological constant, $\Lambda$ \cite{plank-data}.\\
If, as we try to show, the second term in \eqref{0-5} represents the dust matter, it is reasonable
to ask about the implications of $\Lambda$ for the model.\\
The background equations are derived from Eqs. \eqref{2-fri} and \eqref{2-8} as
\begin{subequations}
\begin{align}
        &H^{2}=\frac{\kappa^{2}}{6}\left[\dot{\varphi}^{2}9\alpha^{2}H^{2}+2\Lambda\right]\label{4-1-1},\\
       &H\ddot{\varphi}+\dot{\varphi}(2\dot{H}+3H^{2})=0\label{4-1-2}.
\end{align}
\end{subequations}
It is not possible to find an explicit solution for the Hubble parameter as a function of time from the above equations.
Here useful information can be derived from the following equation:
\begin{equation}\label{4-2}
 \dot{ H}=-3H^{2}\frac{3H^{2}-\kappa^{2}\Lambda}{6H^{2}-\kappa^{2}\Lambda}\equiv f(H),
\end{equation}
which is obtained by taking the time derivative of \eqref{4-1-1}, and using \eqref{4-1-1} and \eqref{4-1-2}. Note
that the denominator of Eq. \eqref{4-2} does not vanish, because, from \eqref{4-1-1} it follows that
\begin{equation}\label{4-3}
 H^{2}\geq\frac{\kappa^{2}\Lambda}{3},
\end{equation}
otherwise $\dot{\varphi}^{2}<0$.\\
Using the above bound, it turns out that Eq. \eqref{4-2} has one fixed point at $H^{2}_{*}=\kappa^{2}\Lambda/3$, which is the de Sitter point . Also,
\begin{equation}\label{4-4}
 \frac{ df(H)}{dH}\mid_{H_{*}}=-6\sqrt{\frac{\kappa^{2}\Lambda}{3}}<0,
\end{equation}
so $H_{*}$ is the stable fixed point. It means that with any initial condition, for Eqs. \eqref{4-1-1} and \eqref{4-1-2}, the Universe evolves to
the accelerated expansion phase, which is consistent with the statements at the beginning of this section.\\
In summary, near the fixed point( de Sitter phase), the cosmological constant term is dominated and the dust term in \eqref{0-5} will be irrelevant. This note
is clear from \eqref{4-1-2} in which at the de Sitter background we have $\dot{\varphi}\propto\exp[-3H_{*}t]$ that leads to $\rho_{d}\propto\exp[-6H_{*}t]$.  Thus, at the late time, the action is reduced to the Hilbert-Einstein action and the cosmological constant term.\\
Hence, to find effects of the second term of \eqref{0-5} with the cosmological constant,
it seems that the other limit, $\kappa^{2}\Lambda\ll H^{2}$, is more interesting.
Physically it means the dust matter dominated universe with the small cosmological constant. For this limit, from \eqref{4-2} we have
\begin{equation}\label{4-5}
 \dot{ H}=-\frac{3}{2}H^{2}(1-\frac{\Lambda\kappa^{2}}{6H^{2}}),
\end{equation}
so
\begin{equation}\label{4-6}
 H=\frac{2}{3t}+\frac{1}{12}\Lambda\kappa^{2}t,\quad  a=t^{\frac{2}{3}}(1+\frac{1}{54}\frac{\Lambda\kappa^{2}}{H^{2}}).
\end{equation}
If the effective density and the effective pressure are defined as $\bar{\rho}_{eff}\equiv3H^{2}/\kappa^{2}$, $\bar{p}_{eff}\equiv(-3H^{2}-2\dot{H})/\kappa^{2}$ respectively, it follows that
\begin{equation}\label{den}
\bar{\rho}_{eff}=\frac{4}{3\kappa^{2}t^{2}}+\frac{\Lambda}{3},\quad\bar{p}_{eff}=-\frac{\Lambda}{2},
\end{equation}
where $\kappa^{2}\Lambda\ll H^{2}$ is applied.
\subsection{\label{sec:level1}	The scalar metric perturbations}
For reasons that will soon become clear, it is worth mentioning some notes about Eq. \eqref{2-p-2}. For the pure dust, $V(\varphi)=0$, the left-
hand side of Eq. \eqref{2-p-2} is vanished, which does not happen for $V(\varphi)\neq0$. Therefore,
it is not trivial that the solutions of the scalar metric perturbations for $V(\varphi)\neq0$ approach to the solutions of $V(\varphi)=0$ if we, simply, take
$V(\varphi)\rightarrow0$ in the solutions. We pointed out that how eventually one can find constraint equations for the pure dust matter from Eq. \eqref{2-p-2},  which are Eqs.\eqref{3-3}, \eqref{3-7} and \eqref{3-8}.\\
Hence, if $\zeta(V)$ is the scalar metric perturbation for $V(\varphi)\neq0$ and
\begin{equation}\label{4-2-1}
\lim_{V\rightarrow0}\zeta(V)\neq\zeta|_{\textit{pure dust matter}},
\end{equation}
it turns out that the solution has not correct behavior at the stated limit. Thus, although such solutions are correct for $V(\varphi)\neq0$, they must be excluded at the stated limit.
Later, we clarify the above statements by an explicit example.\\
For the cosmological constant case, we focus on $\kappa^{2}\Lambda\ll H^{2}$ limit. The constraint equation \eqref{2-p-2} gives
\begin{equation}\label{4-2-2}
-\frac{\Lambda\kappa^{2}}{3H}\partial^{2}\psi=(\frac{2}{3}+\frac{\Lambda\kappa^{2}}{3H^{2}})\partial^{2}\zeta-9Ha^{2}(\frac{1}{3}-\frac{\Lambda\kappa^{2}}{9H^{2}})\dot{\zeta}.
\end{equation}
Dividing by $\Lambda\kappa^{2}/3H$ then inserting the result in \eqref{2-seond-scalar}, to eliminate $\psi$, yields
\begin{equation}\label{4-2-3}
S^{S}_{\Lambda}=\frac{1}{\kappa^{2}}\int dtd^{3}x\left[\frac{1}{9}-\frac{2}{3}\frac{H^{2}}{\Lambda\kappa^{2}}\right]a\zeta\partial^{2}\zeta
-\left[7-12\frac{\Lambda\kappa^{2}}{H^{2}}\right]a^{3}\dot{\zeta}^{2},
\end{equation}
where the stated limit, $\kappa^{2}\Lambda\ll H^{2}$, is used.
By variation with respect to $\zeta$ and using $\kappa^{2}\Lambda\ll H^{2}$, it turns out that
\begin{equation}\label{4-2-4}
\ddot{\zeta}-\left(\frac{5}{4}\frac{\Lambda\kappa^{2}}{H}\right)\dot{\zeta}+\frac{1}{36}\left(\frac{5}{6}\frac{\Lambda\kappa^{2}}{H^{2}}+2\right)\frac{\partial^{2}\zeta}{a^{2}}=0.
\end{equation}
Before we analyze the above equation, let us clarify our statements at the beginning of this subsection by this
explicit example.\\
If we naively take $\Lambda=0$ in Eq. \eqref{4-2-4}, we find
\begin{equation}\label{4-2-5}
\ddot{\zeta}+\frac{1}{18}\frac{\partial^{2}\zeta}{a^{2}}=0,\quad \Lambda\rightarrow0.
\end{equation}
Equation \eqref{4-2-5} has not any solution that approaches to pure dust case, i.e. Eq. \eqref{3-3}. For this case it is easy to find why the procedure
,which gives Eq. \eqref{4-2-5}, is wrong. Recall that after Eq. \eqref{4-2-2}, to eliminate $\psi$, we divided by $\Lambda\kappa^{2}/3H$ and
to obtain Eq. \eqref{4-2-4}, it has been assumed that $\Lambda\neq0$. Obviously the procedure is not correct when $\Lambda\rightarrow0$.\\
Hence, the scalar metric perturbations, $\zeta$, which satisfy Eq. \eqref{4-2-5} must be excluded. In other words, Eq. \eqref{4-2-5} must
be replaced with the correct version, i.e. Eqs.\eqref{3-3}, \eqref{3-7} and \eqref{3-8}.\\
To analyze Eq. \eqref{4-2-4}, let us define the following variables:
\begin{equation}\label{4-2-6}
  K\equiv \exp[\frac{-5}{8}\Lambda\kappa^{2}\int dtH^{-1}],\quad \Xi_{\Lambda}\equiv K\zeta_{k},
\end{equation}
where $H$ is given by \eqref{4-6} and
\begin{equation}\label{4-2-7}
\zeta=\int\frac{d^{3}k}{(2\pi)^{\frac{3}{2}}}\zeta_{k}e^{i\vec{k}.\vec{x}}.
\end{equation}
From Eqs. \eqref{4-2-4}, \eqref{4-2-6} and Eq. \eqref{4-2-7}, we find
\begin{equation}\label{4-2-8}
\ddot{\Xi}-\left[\frac{\ddot{K}}{K}+\frac{k^{2}}{36a^{2}}\left(2+\frac{5}{6}\frac{\Lambda\kappa^{2}}{H^{2}}\right)\right]\Xi=0,
\end{equation}
where
\begin{equation}\label{4-2-9}
\frac{\ddot{K}}{K}=\frac{-5}{16}\Lambda\kappa^{2}\left(3-\frac{7}{4}\frac{\Lambda\kappa^{2}}{H^{2}}\right).
\end{equation}
As for outside the horizon, we take wavelengths which are not only greater than the Hubble length $k/a\ll H$, but also greater than the scale
of cosmological constant $k/a\ll \Lambda^{\frac{1}{2}}\kappa$. For this limit, one of the solutions of Eq. \eqref{4-2-8} is $\Xi\approx K$, so
\begin{equation}\label{4-2-10}
\zeta_{k}|_{out}\approx C_{\Lambda},
\end{equation}
where $C_{\Lambda}$ is a constant. Therefore, as usual, there exists one solution in which $\dot{\zeta}_{k}|_{out}=0$.\\
The other solution for outside the horizon,$^{II}\zeta_{k}|_{out}$, satisfies the following equation:
\begin{equation}\label{4-2-10}
^{II}\dot{\zeta}_{k}|_{out}\approx \exp\left[\frac{5}{4}\Lambda\kappa^{2}\int dtH^{-1}\right].
\end{equation}
Using \eqref{4-6} and $\kappa^{2}\Lambda\ll H^{2}$ it follows that
\begin{equation}\label{moz}
^{II}\zeta_{k}|_{out}=\frac{2}{3H}\left(1+\frac{5}{18}\frac{\Lambda\kappa^{2}}{H^{2}}\right).
\end{equation}
From \eqref{4-6} it is clear that $H$ is increased with time.
So, $^{II}\zeta_{k}|_{out}$ is a decaying solution and we ignore it.\\
Inserting $\zeta_{k}|_{out}$ in Eq. \eqref{4-2-2}, gives $\psi_{k}|_{out}$ as
\begin{equation}\label{si}
\psi_{k}|_{out}=-(1+\frac{2H}{\Lambda\kappa^{2}})\zeta_{k}|_{out}.
\end{equation}
With the same procedures in the previous section, one can obtain, $\delta\rho/\bar{\rho}_{eff}|_{out}$. The Fourier coefficient of this quantity is
\begin{equation}\label{qu}
\frac{\delta\rho_{k}}{\bar{\rho}_{eff}}|_{out}=k^{2}a(1-\frac{1}{6}\frac{\Lambda\kappa^{2}}{H^{2}})\zeta|_{out},
\end{equation}
where the decaying modes have been neglected. Also
\begin{equation}\label{isoa}
\frac{\delta\Pi^{S}_{k}}{\bar{p}_{eff}}|_{out}=\mathcal{O}(a^{-3}),\quad \frac{\delta p_{k}}{\bar{p}_{eff}}|_{out}=\mathcal{O}(a^{-3}).
\end{equation}
Here $\bar{p}_{eff}\neq0$ and is given by \eqref{den}.\\
Returning now to inside the horizon limit, $k/a\gg H\gg\kappa\sqrt{\Lambda}$.
It is rather clear that the very small wavelengths( compare with the horizon, $1/H$, and the cosmological constant scale, $1/\kappa\sqrt{\Lambda}$)
do not``feel" the effects of $\Lambda$. Hence, we expect that their dynamics are determined by equations which were obtained in the previous section.\\
Mathematically, the same result is obtained if we notice how inside the horizon limit is obtained from Eqs. \eqref{4-2-5} and \eqref{4-2-8}. The
stated limit is obtained by dropping terms proportional to $H$ in favor of terms proportional to $k/a$. But our equations, in this subsection, are given under condition that $H\gg\kappa\sqrt{\Lambda}$.
Therefore, by taking inside the horizon limit, we implicitly drop terms in which $\Lambda$ is appear. So, it is not
amazing why Eq. \eqref{4-2-5} is obtained at the end of this procedure. Hence, regarding discussions at the beginning of this subsection and after Eq. \eqref{4-2-5},
to study inside the horizon limit, Eqs. \eqref{3-7} and \eqref{3-8} must be used.
\subsection{\label{sec:level1}	The tensor metric perturbations}
Fortunately, for the tensor metric perturbations we are not confronted with constraint equations. So
the results that will be obtained in this part, can be checked, simply, by taking $\Lambda\rightarrow0$ in the results.\\
Using Eq. \eqref{4-1-1} in \eqref{3-19}, with $x\rightarrow\sqrt{2}x$, then gives
\begin{equation}\label{4-4-1}
S_{\Lambda}^{T}=\frac{\sqrt{2}}{4}\int dtd^{3}x\left[(\frac{2}{3}-\frac{\kappa^{2}\Lambda}{18H^{2}})a\gamma_{ij}\partial^{2}\gamma_{ij}+
(\frac{2}{3}+\frac{\kappa^{2}\Lambda}{9H^{2}})a^{3}\dot{\gamma_{ij}}\dot{\gamma_{ij}}\right].
\end{equation}
Variation with respect to $\gamma_{ij}$ and then inserting Eq.\eqref{1-5} in the result, gives
\begin{equation}\label{4-4-2}
\left[(\frac{2}{3}+\frac{\kappa^{2}\Lambda}{9H^{2}})a^{3}\right]\ddot{\gamma}^{s}_{k}+
\frac{d}{dt}\left[(\frac{2}{3}+\frac{\kappa^{2}\Lambda}{9H^{2}})a^{3}\right]\dot{\gamma}^{s}_{k}
+\left(\frac{2}{3}-\frac{\kappa^{2}\Lambda}{18H^{2}}\right)k^{2}a\gamma^{s}_{k}=0.
\end{equation}
Comparing the above expression with Eq.\eqref{3-22}, it follows that
\begin{eqnarray}\label{4-4-3}
-16\pi Ga^{3}\Pi^{T}_{ij}=&&\left[(\frac{\kappa^{2}\Lambda}{6H^{2}})a^{3}\right]\ddot{\gamma}^{s}_{k}+
\frac{d}{dt}\left[(\frac{\kappa^{2}\Lambda}{6H^{2}})a^{3}\right]\dot{\gamma}^{s}_{k}\nonumber\\
&&-\left(\frac{\kappa^{2}\Lambda}{12H^{2}}\right)k^{2}a\gamma^{s}_{k}.
\end{eqnarray}
Note that for this case $\Pi^{T}_{ij}$ is proportional to $\kappa^{2}\Lambda/H^{2}$, therefor, in the absence of any growing mode
for $\gamma_{k}^{s}$, it will be suppressed. Our aim is to obtain expressions for the outside and inside the horizon of $\gamma^{s}_{k}$, and then find $\Pi^{T}_{ij}$.\\
To study \eqref{4-4-2}, we define the following variables:
\begin{equation}\label{4-4-4}
h\equiv a^{\frac{3}{2}}\left(\frac{2}{3}+\frac{\kappa^{2}\Lambda}{9H^{2}}\right)^{\frac{1}{2}},\quad \Gamma_{\Lambda}^{s}\equiv h\gamma^{s}_{k}.
\end{equation}
Substituting \eqref{4-4-4} into Eq.\eqref{4-4-2} and using $\Lambda\kappa^{2}\ll H^{2}$ yields
\begin{equation}\label{4-4-5}
\ddot{ \Gamma}_{\Lambda}^{s}+\left[\frac{k^{2}}{a^{2}}(1-\frac{\kappa^{2}\Lambda}{4H^{2}})-\frac{\ddot{h}}{h}\right]\Gamma_{\Lambda}^{s}=0,
\end{equation}
where it can be shown that
\begin{equation}\label{4-4-6}
 \frac{\ddot{h}}{h}=\frac{3}{2}\Lambda\kappa^{2}(1-\frac{5}{12}\frac{\kappa^{2}\Lambda}{H^{2}})
\end{equation}
For outside the horizon, $k/a\ll H$, Eq. \eqref{4-4-5} has the following solution:
\begin{equation}\label{4-4-7}
\Gamma_{\Lambda}^{s}|_{out}=C^{\Lambda}_{T}h\Rightarrow \gamma^{s}_{k}|_{out}=C^{\Lambda}_{T},
\end{equation}
where $C^{\Lambda}_{T}$ is a constant. Hence, as usual $\dot{\gamma}^{s}_{k}|_{out}=0$.
The other solution of Eq.\eqref{4-4-2} takes the following form for outside the horizon:
\begin{equation}\label{4-4-8}
\gamma^{s}_{k}|_{out}=\int\frac{dt}{a^{3}\left(\frac{2}{3}+\frac{\kappa^{2}\Lambda}{9H^{2}}\right)}=
\frac{3}{2}\int\frac{dt}{a^{3}}-\frac{1}{4}\int\frac{\kappa^{2}\Lambda dt}{H^{2}a^{3}},
\end{equation}
which is decaying mode.\\
For deep inside the horizon, $k/a\gg H$, the WKB method gives the following solutions for $\Gamma^{s}$:
\begin{equation}\label{4-4-9}
\Gamma^{s}_{\Lambda}|_{in}\approx \frac{\sqrt{a}}{\sqrt[4]{k^{2}(1-\frac{\kappa^{2}\Lambda}{4H^{2}})}}\exp\left[\pm i\int\frac{k}{a}\sqrt{(1-\frac{\kappa^{2}\Lambda}{4H^{2}})}dt\right].
\end{equation}
Hence,
\begin{equation}\label{4-4-10}
\gamma^{s}_{ij}|_{in}\approx \frac{(1-\frac{\kappa^{2}\Lambda}{48H^{2}})}{\sqrt{k}a}\exp\left[\pm i\int\frac{k}{a}(1-\frac{\kappa^{2}\Lambda}{8H^{2}})dt\right],
\end{equation}
which are decaying modes. To check the above results, if $\Lambda\rightarrow0$ and $a\rightarrow t^{2/3}$, the solution approaches to Eq. \eqref{1-7}.\\
Therefore, for this case, from Eqs. \eqref{4-4-3}, \eqref{4-4-7}, \eqref{4-4-8} and \eqref{4-4-10}, it turns out that
\begin{equation}\label{4-4-11}
\Pi^{T}_{ij}|_{out}=\Pi^{T}_{ij}|_{in}=0
\end{equation}
where decaying modes have been neglected.
\section{\label{sec:level1} the	quadratic term }
So far, we have shown that the second term in \eqref{0-5} can be interpreted as the dust matter term. In this section, the quadratic term with a ``mass" parameter, $m$, is considered as $V(\varphi)=\frac{1}{2}m^{2}\varphi^{2}$.
We will give evidence to show how this extension can be ruled out. At first, it must be noted that, in the action \eqref{0-5}, the standard kinetic term does not exist. Therefore, one cannot regard $m$ as the mass of dust matter, which is a rather misleading concept.\\
Let us rewrite the background equations, Eqs. \eqref{2-fri} and \eqref{2-8}, as
\begin{subequations}
\begin{align}
        &H^{2}=\frac{\kappa^{2}}{6}\left[\dot{\varphi}^{2}9\alpha^{2}H^{2}+2V(\varphi)\right]\label{5-1-1},\\
       &\ddot{\varphi}-3H\dot{\varphi}w_{eff}=-\frac{1}{3H^{2}\alpha^{2}}\frac{dV(\varphi)}{d\varphi}\label{5-1-2},
\end{align}
\end{subequations}
where $w_{eff}$ is the effective equation of state
\begin{equation}\label{5-2}
w_{eff}=-1-\frac{2}{3}\frac{\dot{H}}{H^{2}}\quad.
\end{equation}
In Sec. I, we argued that for a scalar field which is minimally coupled to gravity as \eqref{0-1}, eventually the scalar field oscillates around
the minimum of the potential and behaves like the dust matter with $H=2/3t$ ($w_{eff}=0$).\\
Here, we first show that a different scenario arises for the scalar field in \eqref{0-5} when a general potential, which has a minimum, is presented.\\
Note that when $w_{eff}\approx-1$( e.g. during the inflationary phase), the second term on the left-hand side in \eqref{5-1-2} is positive,
and acts as a dissipative force. So, the scalar field rolls toward the minimum of the potential. But at $H=2/3t$, the effective equation of state $w_{eff}$, is vanished and, after some time,
its sign may be changed. So, it is possible that the second term on the left-hand side in \eqref{5-1-2} acts as a driving force. Another point is that the
right-hand side of \eqref{5-1-2} depends on $H$ which, generally, is not constant.
This behavior shows that we need a different analysis for this model.\\
The above statement is supported by explicit calculations for the quadratic potential\cite{ghalee}.
It has been shown that for $V(\varphi)=\frac{1}{2}m^{2}\varphi^{2}$, the following solution for the scalar field is an attractor solution of Eqs. \eqref{5-1-1} and \eqref{5-1-2}\cite{ghalee}:
\begin{equation}\label{5-3}
\varphi(t)=\frac{\sqrt{6}H(t)}{\kappa m}\cos\left[\Omega^{2}t^{2}\right],
\end{equation}
where is viable solution for $m\gg H$ and
\begin{equation}\label{omega}
\Omega^{2}\equiv \frac{m}{2\alpha}(2-\sqrt{2})(\sqrt{2}-\frac{1}{2}).
\end{equation}
Also
\begin{equation}\label{5-4}
H(t)=\frac{2}{3(2-\sqrt{2})t}\Rightarrow w_{eff}\approx-0.41\quad.
\end{equation}
The solutions, \eqref{5-3} and \eqref{5-4}, are very different from \eqref{0-4}.\\
From \eqref{5-4} it turns out that $w_{eff}<-1/3$. So, the Universe is driven to accelerated expansion phase by
the scalar field. But, observations show that at the present time $w_{eff}\approx-1$\cite{plank-data}. Also, we have no
evidence to show a stable phase for the Universe with $w_{eff}\approx-0.41$.\\
So, even at the background level, the quadratic term can be ruled out.
\subsection{\label{sec:level1}	The scalar metric perturbations}
Regarding the last sentence, any motivation to calculate the metric perturbations, for this case, is killed.\\
But here, for model builder who are interested in to extend the model, we show how an analytic solution similar to \eqref{5-3}, helps us
to impose a restriction on the parameters of the model. For our aim, it is sufficient to focus on outside the horizon limit.\\
Using the background equations, Eqs.\eqref{5-1-1}, \eqref{5-3}, \eqref{5-4} and the action for the scalar metric perturbations ,\eqref{Aaction}, the equation of motion for $\zeta$ can be derived.
Outside the horizon limit of the equation becomes
\begin{equation}\label{5-2-1}
\begin{split}
&\frac{d}{dt}\left[a^3\left(3\sin^{2}\left[\Omega^{2}t^{2}\right]\cos^{2}\left[\Omega^{2}t^{2}\right]+4\sin^{2}\left[\Omega^{2}t^{2}\right]\right)\right]\dot{\zeta}_{out} \cr
&+\left[a^3\left(3\sin^{2}\left[\Omega^{2}t^{2}\right]\cos^{2}\left[\Omega^{2}t^{2}\right]+4\sin^{2}\left[\Omega^{2}t^{2}\right]\right)\right]\ddot{\zeta}_{out} \cr
&+4\Omega^{2}t\tan\left[\Omega^{2}t^{2}\right]\dot{\zeta}_{out}=0.
\end{split}
\end{equation}
One of the solutions, as usual, is $\zeta_{out}=C_{\varphi}$, where $C_{\varphi}$ is a constant. The other solution decays as $\int a^{-3}dt$.\\
As the previous cases, $\delta\rho/\bar{\rho}$ can be calculated. The result is
 \begin{equation}\label{send}
 \frac{\delta\rho}{\bar{\rho}}|_{out}=\frac{2}{9}\frac{k^{2}}{a^{2}H^{2}}\tan^{2}\left[\Omega^{2}t^{2}\right]\zeta_{out}.
 \end{equation}
So, to have a viable limit, we must take $\Omega^{2}t^{2}\approx m/(\alpha H^{2})\ll1$. Since the solution $\eqref{5-3}$
is obtained under the condition that $m\gg H$, it turns out that we must take $\alpha H\gg1$.
\subsection{\label{sec:level1}	The tensor metric perturbations}
Using Eq. \eqref{5-1-1} in \eqref{3-19} (with $x\rightarrow\sqrt{2}x$) for the quadratic term, we have
\begin{eqnarray}\label{5-3-1}
S_{\varphi}^{T}=\frac{\sqrt{2}}{6}\int dtd^{3}x&&[(1-\frac{\kappa^{2}m^{2}\varphi^{2}}{24H^{2}})a\gamma_{ij}\partial^{2}\gamma_{ij}\nonumber\\
&&+(1+\frac{\kappa^{2}m^{2}\varphi^{2}}{12H^{2}})a^{3}\dot{\gamma_{ij}}\dot{\gamma_{ij}}],
\end{eqnarray}
where $\varphi$ is given by \eqref{5-3}. Using Eqs. \eqref{1-5}, \eqref{5-3}, and the above action, the equation of motion for $\gamma^{s}_{k}$ is obtained as
\begin{eqnarray}\label{5-3-2}
&&\left[(1+\frac{\cos^{2}[\Omega^{2}t^{2}]}{2})a^{3}\right]\ddot{\gamma}^{s}_{k}+\frac{d}{dt}\left[(1+\frac{\cos^{2}[\Omega^{2}t^{2}]}{2})a^{3}\right]\dot{\gamma}^{s}_{k}\nonumber\\
&&+\left(1-\frac{\cos^{2}[\Omega^{2}t^{2}]}{4}\right)k^{2}a\gamma^{s}_{k}=0.
\end{eqnarray}
So, from Eq.\eqref{3-22} the tensor part of anisotropic stress can be read as
\begin{eqnarray}\label{5-3-3}
-16\pi Ga^{3}\Pi^{T}_{ij}=&&\left[\frac{\cos^{2}[\Omega^{2}t^{2}]}{2}a^{3}\right]\ddot{\gamma}^{s}_{k}+\frac{d}{dt}\left[(\frac{\cos^{2}[\Omega^{2}t^{2}]}{2})a^{3}\right]\dot{\gamma}^{s}_{k}\nonumber\\
&&-\frac{\cos^{2}[\Omega^{2}t^{2}]}{4}k^{2}a\gamma^{s}_{k}.
\end{eqnarray}
As the previous case, to obtain expressions for $\Pi^{T}_{ij}$, we first seek solutions for outside and inside the horizon of $\gamma_{s}^{k}$.\\
Using the following variables,
\begin{equation}\label{5-3-4}
z\equiv\left(1+\frac{\cos^{2}[\Omega^{2}t^{2}]}{2}\right)^{\frac{1}{2}}a^{3/2},\quad \Gamma^{s}_{\varphi}\equiv z\gamma^{s}_{k},
\end{equation}
Eq. \eqref{5-3-2} takes the following form:
\begin{equation}\label{5-3-5}
\ddot{\Gamma^{s}_{\varphi}}+\left[\frac{k^{2}}{a^2}\frac{4-\cos^{2}[\Omega^{2}t^{2}]}{4+2\cos^{2}[\Omega^{2}t^{2}]}-\frac{\ddot{z}}{z}\right]\Gamma^{s}_{\varphi}=0.
\end{equation}
So, for outside the horizon, $k/a\ll H$, one of the solutions is
\begin{equation}\label{5-3-6}
\Gamma^{s}_{\varphi}|_{out}=C^{\varphi}_{T}z,\Rightarrow\gamma_{k}^{s}|_{out}=C_{\varphi}^{T},
\end{equation}
where $C_{\varphi}^{T}$ is a constant. The other solution is
\begin{equation}\label{5-3-7}
 \gamma_{k}^{s}|_{out}=\int\frac{2dt}{a^3(2+\cos^{2}[\Omega^{2}t^{2}])}
\end{equation}
which is decaying mode.\\
As for inside the horizon,$k/a\gg H$, the WKB method gives two solutions for $\Gamma^{s}_{\varphi}$, and from \eqref{5-3-4} the following solutions for $\gamma_{k}^{s}$ are obtained:
\begin{equation}\label{5-3-8}
\gamma_{k}^{s}|_{in}\approx \frac{1}{a\sqrt{f(t)}}\exp \left[\pm i\int\frac{k}{a}\sqrt{\frac{4-\cos^{2}[\Omega^{2}t^{2}]}{4+2\cos^{2}[\Omega^{2}t^{2}]}}dt\right],
\end{equation}
where
\begin{equation}\label{5-3-9}
  f(t)\equiv 16+4\cos^{2}[\Omega^{2}t^{2}]-2\cos^{4}[\Omega^{2}t^{2}].
\end{equation}
Both solutions for inside the horizon are decaying mode.\\
Hence, from the above statements about the solutions of $\gamma_{k}^{s}$, and Eq. \eqref{5-3-3}, it turns out that
\begin{equation}\label{5-3-10}
\Pi_{ij}^{T}|_{in}=\Pi_{ij}^{T}|_{out}=0.
\end{equation}
\section{\label{sec:level1}	Summary and discussion}
The dust matter is one of the main substances of the Universe. The standard textbooks, e.g. \cite{mukhanov,weinberg},
implicitly use the intuitive picture borrowed from the Newtonian cosmology , and define it by the stress-energy tensor as pressureless matter, $\bar{p}=0$.\\
In this work, we provide an alternative approach. We have shown that it can be possible to interpret the second
term in \eqref{0-5} as the dust matter term. To show this point, the background and perturbative equations are studied.\\
For the pure dust matter, we have found that $\delta\equiv\delta\rho/\overline{\rho}$ and $\delta\Pi_{ij}^{T}$
have the same behaviors as are found in the textbooks, e.g.\cite{mukhanov}, for the dust matter. Furthermore, the expressions for $\delta p$ and $\delta\Pi^{S}$
are decaying terms.\\
We have also considered the dynamics of the model with the cosmological constant. We
have shown that the dynamics of our model, both for the dust matter dominated era with the small
cosmological constant and the late time behavior, are consistent with the observation\cite{plank-data}.\\
Therefore the model has not any exotic behavior.\\
If the model can be interpreted as the dust matter, an inevitable question arises as follows: is it
possible to obtain new predictions from it and how is possible to extend it? Regarding the question, an implicit
attempt is made by extending the model with inclusion of the quadratic term in the action.
It turns out that this version is inconsistent with the observations. But the attempt has two
lessons for us. First, we have shown that generally a potential term, which has one minimum, results in
a universe with $-1<w_{eff}<0$. Thus, it seems we can rule them out. The second lesson is that, although
the kinetic term is nonminimally coupled to the tensor field, we have no problem with
the gravitational waves and any constraint on the parameters of the model- in the extend version- comes from
the scalar metric perturbations.
\begin{acknowledgments}
I am grateful for helpful discussions with F. Arash.
\end{acknowledgments}
\appendix
\section{}
In this paper, the Hubble parameter is obtained as a function of time. Therefore to write the second order action for the scalar metric perturbations
the terms in which $\dot{H}$ appears are kept. The result for $V(\varphi)\neq0$ is
\begin{equation}\label{Aaction}
\begin{split}
S^{S}=\frac{1}{2\kappa^{2}}\int dtd^{3}x&\left[6\frac{\dot{H}}{H^{2}}\left(\frac{(2-\alpha^{2}\kappa^{2}\dot{\varphi}^{2})^{2}}{(3\alpha^{2}\kappa^{2}\dot{\varphi}^{2}-2)^{2}}\right)\right]a\alpha^{2}\kappa^{2}\dot{\varphi}^{2}\zeta\partial^{2}\zeta \cr
+&\left[\frac{12+\alpha^{2}\kappa^{2}\dot{\varphi}^{2}(20-21\alpha^{2}\kappa^{2}\dot{\varphi}^{2})}{(3\alpha^{2}\kappa^{2}\dot{\varphi}^{2}-2)^{2}}\right]a\alpha^{2}\kappa^{2}\dot{\varphi}^{2}\zeta\partial^{2}\zeta \cr
-&3\left[\frac{9\alpha^{2}\kappa^{2}\dot{\varphi}^{2}+2}{(3\alpha^{2}\kappa^{2}\dot{\varphi}^{2}-2)}\right]\alpha^{2}\kappa^{2}\dot{\varphi}^{2}a^{3}\dot{\zeta}^{2}.
\end{split}
\end{equation}
\bibliography{apssamp}

\end{document}